\documentclass{webofc}

\usepackage[varg]{txfonts}   
\usepackage{hyperref}
\usepackage{url}
\usepackage{color}
\usepackage{tablefootnote} 
\usepackage{threeparttable}
\hypersetup{colorlinks=true,citecolor=blue,urlcolor=blue,linkcolor=blue}
\begin{document}
\title{Application of Particle Transformer to quark flavor tagging in the ILC project}
%
%

\author{\firstname{Risako} \lastname{Tagami}\inst{1}\fnsep
\and
        \firstname{Taikan} \lastname{Suehara}
             \and
        \firstname{Masaya} \lastname{Ishino}
}

\institute{rtagami@icepp.s.u-tokyo.ac.jp
          }

\abstract{
International Linear Collider (ILC) is a next-generation $e^+e^-$ linear collider to explore Beyond-Standard-Models by precise measurements of Higgs bosons. Jet flavor tagging plays a vital role in the ILC project by identification of $H\to b\bar{b},\,c\bar{c},\,gg,\,s\bar{s}$ to measure Higgs coupling constants and of $HH\to b\bar{b}b\bar{b}$ and $b\bar{b}WW$ which are the main channels to measure the Higgs self-coupling constant.

Jet flavor tagging relies on a large amount of jet information such as particle momenta, energies, and impact parameters, obtained from trajectories of particles within a jet. Since jet flavor tagging is a classification task based on massive amounts of information, machine learning techniques have been utilized for faster and more efficient analysis for the last several decades.

Particle Transformer (ParT) is a machine learning model based on Transformer architecture developed for jet analysis, including jet flavor tagging. In this study, we apply ParT to ILD full simulation data to improve the efficiency of jet flavor tagging.

Our research focused on evaluating the performance of ParT compared to that of the previously used flavor tagging software, LCFIPlus. We will also report the status of the performance of strange tagging on ILD full simulation dataset using ParT, which can be applied to the analysis of Higgs-strange coupling.

}
\maketitle

\def\thefootnote{\textcolor{white}{\fnsymbol{footnote}}}
\section{Introduction}
\label{intro}

After\footnote{This work was carried out in the framework of the ILD Concept Group.} the discovery of the Higgs boson, the search for beyond the Standard Model through precise measurement of Higgs has been expected. There is a consensus among particle physicists to build Higgs Factories as the next-generation accelerator. There are several $\mathrm{e^+e^-}$ Higgs Factories currently under consideration. ILC project is one of the Higgs Factory projects.  

With more precise measurements of Higgs, the effects of SUSY and other new TeV physics models can be seen. For instance, if there is a new physics phenomenon at 1 TeV, the offset from the Standard Model will be about 6$\,\%$ so that accuracy of about 1$\,\%$ will be required. In the ILC project, the accuracy of $H\to b\bar{b}$ measurements is expected to be $\sim 1\%$. For precise measurements of Higgs, especially Higgs coupling constants of $H\to b\bar{b},\,c\bar{c},\,gg,\,s\bar{s}$, the performance of flavor tagging should be improved.

Jet flavor tagging is a classification task using the information of each jet such as impact parameters of tracks and number of vertices. Bottom jets have two secondary vertices, whereas charm jets have one secondary vertex. Strange jets have more strange hadron like Kaon, whereas light jets have less. Those differences between jets should be detected by jet flavor tagging.  

In ILC or CLIC studies, LCFIPlus\cite{Suehara_2016} has been used for flavor tagging. This is a flavor tagging software using machine learning technique of boosted decision tree (BDT) classification. Recently, a research group in CERN presented a new model for flavor tagging called ParticleNet\cite{Qu_2020} using Graph Neural Network (GNN) model. As below, ParticleNet achieved better performance\cite{FCCPN} than LCFIPlus, although there are differences between them in terms of the datasets used and their detectors (ILC and FCC-ee).

\begin{table}[h]
    \centering
    \caption{The comparison of the performance of $b$-jet flavor tagging between using LCFIPlus for ILD full simulation dataset and using ParticleNet for FCC-ee fast simulation dataset.}
    \begin{tabular}{|c||c|c|}
    \hline
    \multicolumn{1}{|c||}{} & \multicolumn{2}{|c|}{$b$-tag 80$\%$ efficiency} \\ \hline
    Method & $c$-bkg acceptance & $uds$-bkg acceptance \\ \hline
    LCFIPlus & $6.3\,\%$ & $0.79\,\%$ \\ \hline
    ParticleNet & $0.40\,\%$ & $<0.01\,\%$ \\ \hline
    \end{tabular}
    
    \label{LCFIPlusandPN}
\end{table}

Lately, Particle Transformer (ParT)\cite{qu2024particletransformerjettagging} has been applied to LHC fast simulation data. The performance of event tagging of fast simulation dataset (JetClass\cite{qu2024particletransformerjettagging}) has been greatly improved by using ParT. In this study, we apply ParT to ILD full simulation dataset to aim at improving the performance of flavor tagging.



\section{Flavor tagging of ILD full simulation dataset using ParT}

\subsection{Dataset}

\label{dataset}

In this study, the full detector simulation events for International Large Detector (ILD)\cite{theildcollaboration2020internationallargedetectorinterim}, one of the ILC detector concepts, are utilized. The ILD has trackers and calorimeters. The trackers consist of silicon vertex detector, silicon inner and outer tracker and main Time Projection Chamber (TPC) at barrel region, while silicon-only tracking at forward region. The calorimeters consist of electromagnetic calorimeter which is composed of tungsten absorber and highly-granular silicon pad sensors with cell size of $5\times 5\,\mathrm{mm^2}$, and hadron calorimeter which is composed of steel absorber and scintillator tile sensors with $3\times 3\,\mathrm{cm^2}$ cells. Endcap and forward calorimeters are also highly segmented with similar configurations. Outside the hadron calorimeter, superconducting solenoid with 3.5 Tesla magnetic field and muon detector as well as return yoke are equipped in this order.

The software framework (iLCSoft) consists of Geant4 based detector simulation with the ILD detector setup (DDsim), digitizer of trackers and calorimeter hits emulating detector effects and a tracking software as low-level reconstruction. For the reconstruction of particles and jets, standard Particle Flow Algorithm (PandoraPFA) and Durham jet clustering algorithm are utilized, respectively. Particles inside each jet are used to produce input variables for the flavor tagging network.

The samples used in this study are $\mathrm{e}^+\mathrm{e}^-\to q\bar{q}$ events at 91 GeV center-of-mass energy (fixed energy), 1M jets in total and $\mathrm{e}^+\mathrm{e}^-\to ZH\to\nu\bar{\nu} q\bar{q}$  at 250 GeV (variable energy, their average is $\sim 60$ GeV), 1M jets in total. 

They are divided into training samples (85$\%$), validation samples (5$\%$) and test samples (15$\%$).

\subsection{Model}
\label{model}

\begin{figure}[h]
\centering
\includegraphics[width=8.5cm,clip]{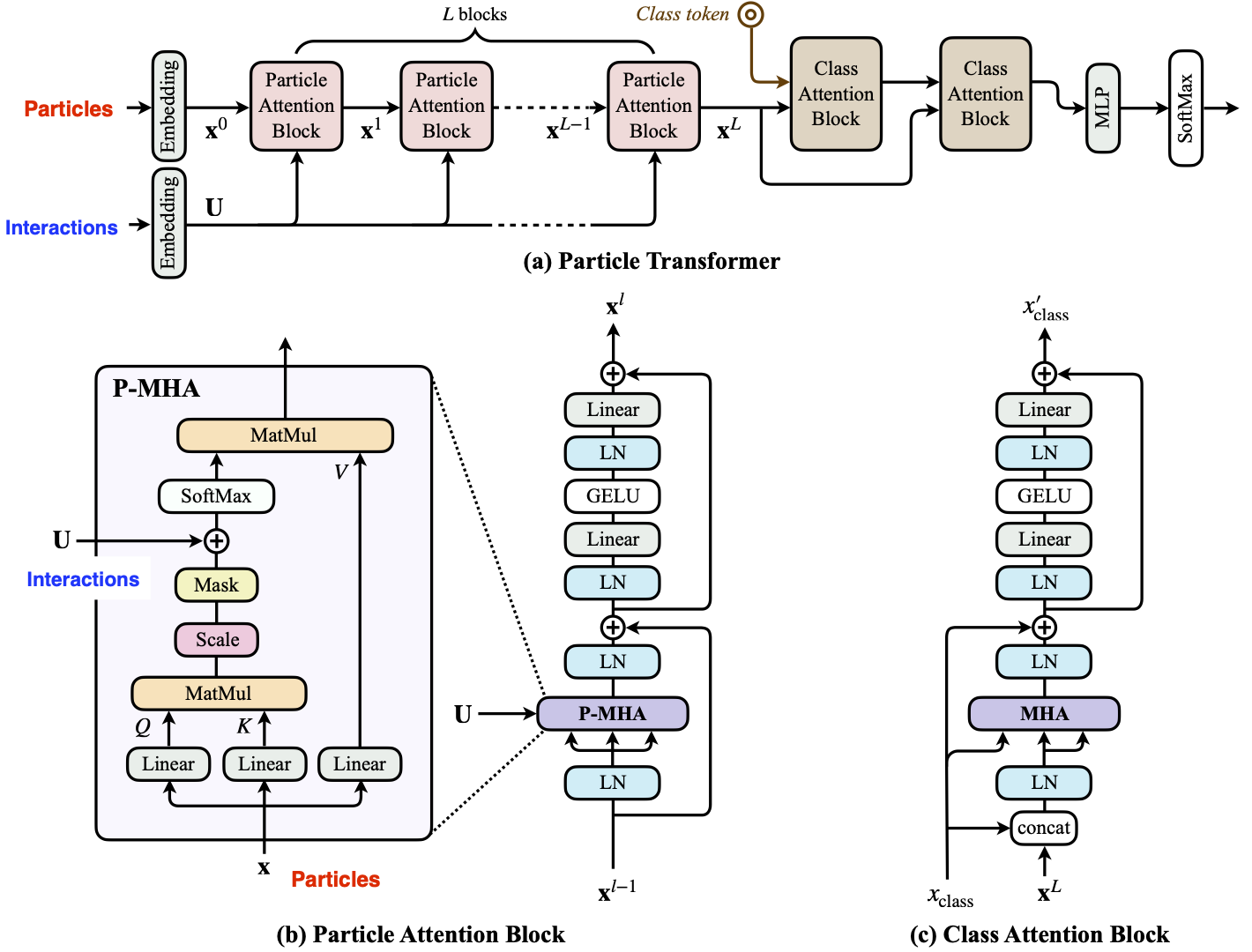}
\caption{The architecture of Particle Transformer's model. The original image is from \cite{qu2024particletransformerjettagging}.}
\label{fig-1}
\end{figure}

The architecture of ParT is as Figure \ref{fig-1}.

The basic structure of ParT is same as Transformer. The difference between standard Transformer and ParT is that ParT has "interactions." Several physical variables are calculated by forming pairs of tracks within jets using their four-momenta and added as bias in the middle of self-attention. 

\subsection{Input variables}
\label{input}

Input variables (features and interactions) are listed in Table \ref{tab:features} and Table \ref{tab:interaction}.

\hspace{10pt}

\def\thefootnote{\fnsymbol{footnote}}

\begin{table}[h]
    \centering
    \caption{Input variables (features).}
    \begin{threeparttable}
    \begin{tabular}{ccc}
    \hline
       Category & Variable & definition \\ \hline
       Impact Parameter& $d_0$ & transverse impact parameter value \\
        & $z_0$ & longitudinal impact parameter value \\
        & ip2D & 2D impact parameter value \\
        & ip3D & 3D impact parameter value \\
       Jet Distance & $d_j$ & \begin{tabular}{cl}
           displacement of tracks from the line \\
           passing interaction point with direction of their jets
       \end{tabular}\\
       Particle ID & Muon & if the particle is a muon ($|\text{pid}|==13$) \\
        & Electron & if the particle is an electron ($|\text{pid}|==11$) \\
        & Gamma & if the particle is a gamma ($|\text{pid}|==22$) \\
        & Charged Hadron & if the particle is a charged hadron \\
        & Neutral Hadron & if the particle is a neutral hadron \\
        & type & PDG ID reconstructed by each Particle ID \\
       Kinematic & Charge & electric charge of the particle \\
        & $\log{\frac{E}{E_{\text{jet}}}}$ & logarithm of the particle's energy relative to the jet energy \\
        & $\Delta\theta$ & $\theta_{\text{track}}-\theta_{\text{jet}}$ \\
        & $\Delta\phi$ & $\phi_{\text{track}}-\phi_{\text{jet}}$ \\
       Track Errors & $\sigma$ & each element of covariant matrix \\ \hline
    \end{tabular}
    \begin{tablenotes}
        \item[*] Impact parameter, Jet Distance and Track Errors are only valid for charged particles. They are set to be -9 for neutral particles to fill in the input data.
    \end{tablenotes}
    \label{tab:features}
    \end{threeparttable}
\end{table}

\begin{table}[h]
    \centering
    \caption{Input variables (interactions). $i$ and $j$ each represent one of a pair of particles.}
    \begin{tabular}{cc}
    \hline
        Variable & definition \\ \hline
        $\log\left(\Delta R\right)$ & logarithm of $\sqrt{(\eta_i-\eta_j)^2+((\phi_i-\phi_j+\pi)\%(2\pi)-\pi)^2}$ \\
        $\log\left(kt\right)$ & logarithm of $\sin\theta_{ij}*p_{\text{min}}$ ($p_{\text{min}}=\text{min}(p_i,\,p_j)$) \\
        $\log\left(z\right)$ & logarithm of $\frac{p^T_{\text{min}}}{p^t_i+p^T_j}$ ($p^T_{\text{min}}=\text{min}(p^T_i,\,p^T_j)$)\\
        $\log\left(\text{inv.mass}\right)$ & logarithm of invariant mass \\ \hline
    \end{tabular}
    \label{tab:interaction}
\end{table}

\subsection{Result}

Performance of three category ($b,\,c,\,d$) flavor tagging on ILD full simulation dataset (at 91 GeV) using ParT is evaluated and compared to the performance using LCFIPlus, and the results are summarized in Table \ref{tab:result-ParT}.

The ParT model is modified to adjust to the ILD full simulation dataset. For instance, features are separated into tracks and neutrals before embedding (linear, Gaussian Error Linear Unit (GELU) activation function) and combined before self-attention block, and several variables of features are standardized by sigmoid function before training. The performance on ILD full simulation dataset (at 250 GeV) with those modification is also listed on the Table \ref{tab:result-ParT}.

\begin{table}[h]
    \centering
    \caption{The performances of $b$-tagging and $c$-tagging. In the top row and the bottom of the table, ILD full simulation dataset (at 250 GeV, 1M jets) is used, while in the middle row, ILD full simulation dataset (at 91 GeV, 1M jets) is used.}
    \begin{tabular}{|c||c|c|c|c|}
    \hline
         & \multicolumn{2}{|c|}{$b$-tag 80$\%$ efficiency} & \multicolumn{2}{|c|}{$c$-tag 50$\%$ efficiency} \\ \hline
        Method & \begin{tabular}{l}
             $c$-bkg \\
             acceptance
        \end{tabular} & \begin{tabular}{l}
             $d$-bkg \\
             acceptance
        \end{tabular} & \begin{tabular}{l}
             $b$-bkg \\
             acceptance
        \end{tabular} & \begin{tabular}{l}
             $d$-bkg \\
             acceptance
        \end{tabular} \\ \hline
        LCFIPlus (at 250 GeV) & $6.3\,\%$ & $0.79\,\%$ & $7.4\,\%$ & $1.2\,\%$ \\ \hline
        ParT (at 91 GeV) & $1.3\,\%$ & $0.25\,\%$ & $1.0\,\%$ & $0.43\,\%$ \\ \hline
        ParT (at 250 GeV, modified) & $0.48\,\%$ & $0.14\,\%$ & $0.86\,\%$ & $0.34\,\%$ \\ \hline
    \end{tabular}
    
    \label{tab:result-ParT}
\end{table}

The performance of $b$-tagging and $c$-tagging is greatly improved by using ParT, about 4.8 times in $c$-background acceptance of $b$-tagging. We also achieved a $\sim 2.7$ times improvement in performance with the modifications. It was found that ParT is also effective for ILD full simulation dataset.

\section{Strange tagging}

\label{strange}

We also report current status of strange jet tagging of ILD full simulation dataset using ParT. 

There are only slight differences in kinematics between strange jet and light jets. Their main difference is the ratio of each particle in jets shown in Table \ref{tab:particle_ratio}. In strange jets, there are more strange hadrons like Kaon.

\begin{table}[h]
    \centering
    \caption{The ratios of track particles with momenta bigger than 5 GeV.}
    \begin{tabular}{|c||c|c|c|}
    \hline
      Particle  & $H\to s\bar{s}$ &  $H\to gg$ & $H\to d\bar{d}$ \\ \hline
      Kaon & 27.1$\,\%$ & 16.2$\,\%$ & 10.7$\,\%$ \\ \hline
      Pion & 63.1$\,\%$ & 69.2$\,\%$ & 79.2$\,\%$ \\ \hline
      Proton & 7.2$\,\%$ & 11.5$\,\%$ & 7.1$\,\%$ \\ \hline
      Electron & 2.1$\,\%$ & 2.3$\,\%$ & 2.5$\,\%$ \\ \hline
      Muon & 0.4$\,\%$ & 0.9$\,\%$ & 0.4$\,\%$ \\ \hline
    \end{tabular}
    \label{tab:particle_ratio}
\end{table}

It is common to use particle identification (ID) of particles in jets for tagging strange jets. We use a new particle ID module called Comprehensive PID (CPID)\cite{einhaus2023cpidcomprehensiveparticleidentification} to improve the performance of strange tagging. With CPID, particles are split by momentum range of 1-100 GeV into 12 momentum bins and their IDs are determined by BDT with the information of reconstructed particle including time of flights and $dE/dx$. As a result, we can obtain particle's probabilities for each species ; kaon, pion, proton, electron and muon. The accuracy of particle ID obtained by CPID is as Table \ref{tab:CPIDaccuracy}. The dataset used is ILD full simulation dataset at 250 GeV with 1M jets.

\begin{table}[h]
    \centering
    \caption{Accuracy rate of CPID in $H\to s\bar{s}$ events with momenta bigger than 5 GeV. Accuracy is defined by the proportion of tracks that CPID (made into 0 or 1 by setting the probability of the Kaon, Pion, Proton, Electron and Muon such that the highest value is 1, and the others are 0) and truth match.}
    \begin{tabular}{|c|c|}
    \hline
       Particle  & Accuracy \\ \hline
       Kaon & 0.74 \\ \hline
       Pion & 0.89 \\ \hline
       Proton & 0.76 \\ \hline
       Electron & 0.38 \\ \hline
       Muon & 0.40 \\ \hline
    \end{tabular}
    \label{tab:CPIDaccuracy}
\end{table}

\begin{figure}[h]
\centering
\includegraphics[width=8cm,clip]{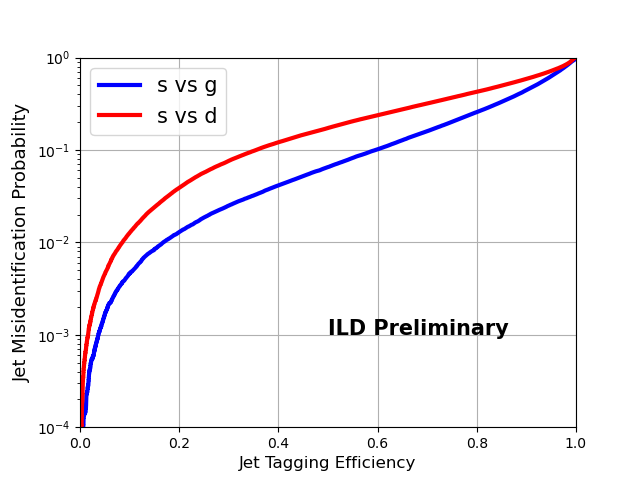}
\caption{The performances of strange tagging of ILD full simulation dataset (at 250 GeV, 1M jets) using ParT. Six category ($b,\,c,\,d,\,g,\,u,\,s$) flavor tagging was performed. Blue line is $g$-background acceptance, and red line is $d$-background acceptance.}
\label{fig-10}       
\end{figure}

We use CPID scores of those five particles and also usual inputs in Table \ref{tab:features} and Table \ref{tab:interaction} for strange tagging. The current performance of strange tagging is as Figure \ref{fig-10}.  

The $g$-background acceptance in $s$-tagging $80\%$ efficiency is $25.7\%$ and the $d$-background acceptance in $s$-tagging $80\%$ efficiency is $42.7\%$. The performance of strange tagging on ILD full simulation dataset is lower than that on FCC-ee dataset, but they are different in terms of their detectors and fast/full simulation. The difference of performance is currently under investigation.

\section{Discussion and Conclusion}
\label{conclusion}

Particle Transformer is also effective for $b,\, c$ jet flavor tagging on ILD full simulation dataset. The performance of $b$-tagging ($c$-background) with ParT is about five times better than with LCFIPlus, and $c$-background acceptance is 0.48$\%$ with $b$-tagging $80\%$ efficiency.

Strange jet tagging with ParT using Comprehensive PID is also performed. The performance of $s$-tagging $80\%$ efficiency ($g$-background) is about $25\%$. We will try to improve the performance, using CPID for particles with their momenta less than 1 GeV.

\section{Acknowledgements}

We would like to thank the LCC generator working group and the ILD software working group for providing the simulation and reconstruction tools and producing the Monte Carlo samples used in this study. This work has benefited from computing services provided by the ILC Virtual Organization, supported by the national resource providers of the EGI Federation and the Open Science GRID.

\end{document}